\documentclass[preprint,showpacs,preprintnumbers,amsmath,amssymb,superscriptaddress]{revtex4}
\include{ams}
\usepackage{amsmath,amssymb,amsthm}
\usepackage{dcolumn}
\usepackage{bm}
\usepackage{graphicx}

\begin{document}

\title{From the currency rate quotations onto strings and brane world scenarios}

\author{D. Horv\'ath}\email{horvath@sors.com}
\author{R. Pincak}\email{pincak@sors.com}
\affiliation{SORS Research a.s, 040 01 Kosice, Slovak Republic}


\date{\today}

\begin{abstract}

In the paper, we study numerically the projections of the real
exchange rate dynamics onto the string-like topology. Our approach
is inspired by the contemporary movements in the string theory. The
string map of data is defined here by the boundary conditions,
characteristic length, real valued and the method of redistribution
of information.  As a practical matter, this map represents the
detrending and data standardization procedure.  We introduced maps
onto 1-end-point and 2-end-point open strings that satisfy the
Dirichlet and Neumann boundary conditions. The questions of the
choice of extra-dimensions, symmetries, duality and ways to the
partial compactification are discussed.  Subsequently, we pass to
higher dimensional and more complex objects. The 2D-Brane was
suggested which incorporated bid-ask spreads. Polarization by the
spread was considered which admitted analyzing arbitrage
opportunities on the market where transaction costs are taken into
account.  The model of the rotating string which naturally yields
calculation of angular momentum is suitable for tracking of several
currency pairs. The systematic way which allows one suggest more
structured maps suitable for a simultaneous  study of several
currency pairs was analyzed by means of the G\^{a}teaux generalized
differential calculus. The effect of the string and brane maps on
test data was studied by comparing their mean statistical
characteristics. The study revealed notable differences between
topologies.  We review the dependence on the characteristic string
length, mean fluctuations and properties of the intra-string
statistics. The study explores the coupling of the string amplitude
and volatility.

\end{abstract}

\maketitle

\section{Introduction}

We are currently in the process of transfer of modern physical ideas
into the neighboring field called econophysics.  The physical
statistical view point has proved fruitful, namely, in the
description of systems where many-body effects dominate. However,
standard, accepted by physicists, bottom-up approaches are
cumbersome or outright impossible to follow the behavior of the
complex economic systems, where autonomous models encounter the
intrinsic variability.

Digital economy is founded on data. In the paper, we suggest and
analyze statistical properties of heuristics based on the currency
rate data which are arranged to mimic the topology of the basic
variants of the physical strings and branes. Our primary motivation
comes from the actual physical concepts~\cite{Mahon2009,Zwie2009};
however, our realization differs from the original attempts in
various significant details. The second aspect of our method is that
it enables a transformation into a format which is useful for an
analysis of a partial trend or relative fluctuations on the time
scale window of interest.

As with most science problems, the representation of data is the key
to efficient and effective solutions.  The underlying link between
our approach and the string theory may be seen in the switching from
a local to a non-local form of the data description. This line
passes from the single price to the multivalued collection of prices
from the temporal neighborhood which we term here the string map. As
we will see later, an important role in our considerations is played
by the distance measure of the string maps. The idea of exploring
the relationship between more intuitive geometric methods and
financial data is not new. The discipline called the geometric data
analysis includes many diverse examples of the conceptual schemes
and theories grounded on the geometric representation and properties
of data. Among them we can emphasize the tree network topology that
exhibits usefulness in the studies of the world-trade
network~\cite{He2010} and other network structures of the market
constructed by means of inter-asset
correlations~\cite{Jung2008,Eryigit2009}. The multivariate
statistical method called {\em cluster} {\em analysis} deals with
data mapping onto representative subsets called {\em
clusters}~\cite{Krish1982}. Here we work on the concept that is
based on projection data into higher dimensional vectors in the
sense of the work~\cite{Grass1983,Ding2010}.  Also, arguments based
on the metrics are consistent with our efforts  but not too obvious
points in common with the original objectives of the nonlinear
analysis.

The string theory development over the past 25 years achieved a high
degree of popularity among physicists~\cite{Polchinski}. The reason
lies in its inherent ability to unify theories that come from
diverse physical spheres. The prime instrument of the unification
represents the concept of extra dimension.  The side-product of
theoretical efforts can be seen in the elimination of the
ultraviolet divergences of Feynman diagrams. However, despite the
considerable achievements, there is a lack of the experimental
verification of the original string theory. In contrast, in the
present work we exploit time-series which can build the family of
the string motivated models of boundary-respecting maps. In a narrow
sense, the purpose of the present data-driven study is to develop
statistical techniques for the analysis of these objects.

The work is organized as follows:  In sec.\ref{sec:Data},  we
specify the data selection and data pretreatment.  In
sec.\ref{sec:d1_map}, we introduce the notion of the string map of
data time series. The symmetry of the maps is discussed in
sec.\ref{sec:sym}. To examine generality and specificity of these
ideas, the calculations have been performed for several
representative currencies and ask (buyer initiated) or bid (seller
initiated) prices. In sec.\ref{sec:Brane}, we give a generalization
for a higher dimensional case which encodes the ask-bid spread
difference. In this section, we also discuss the problem of partial
compactification (subsection \ref{sec:partial}). The statistical
picture (in sections \ref{sec:stat}, \ref{sec:intrastring} and
\ref{sec:polar}) is of primary interest. The analysis of
fluctuations measured in the middle of a string by means of the
statistical moments is carried out in sec.(\ref{sec:distrib}). In
sec.\ref{sec:intrastring}, we focus on the statistics of
intra-string degrees of freedom.  The efforts are in particular
justified by the analytical studies in subsections sec.\ref{sec:exp}
and sec.\ref{sec:per}. They includes a decomposition of internal
string states into Fourier modes in sec.\ref{sec:DFT}. In
sec.\ref{sec:polar}, we will see how the arbitrage opportunities,
including spread, may be characterized by means of the model of {\em
polarized string}. In the same section, we briefly discuss a
reformulation of the popular concept of the correlation sum in terms
of the strings and branes. There are also indications of a synergy
between the string amplitudes and volatility~in sec.\ref{sec:volat}.
Efforts have been made to study inter-currency relations by means of
the projections onto {\em rotating strings}
(sec.\ref{sec:rotstring}) and via a generalization of the derivative
(sec.\ref{sec:Gateaux}).

\section{Data analysis}\label{sec:Data}

First of all we need to mention some facts about data streams we
analyzed. We analyze tick by tick data of EUR/USD, GBP/USD, USD/JPY,
USD/CAD, USD/CHF major currency pairs from the OANDA market maker.
We focussed on the three month period within three selected periods
of 2009 which capture moments of the financial crisis. The streams
are collected in such a way that each stream begins with Monday.
More precisely, we selected periods denoted as {\rm Aug-Sep} (from
August 3rd. to September 7th.), Sep-Oct (Sep.7-Oct.5) and Oct-Nov
(Sep.5-Nov.2).  At first, the data sample has been decimated - only
each 10th tick was considered. This delimits results to the scales
larger than 10 ticks. The mean time corresponding to the string
length $l_{\rm s}$ in ticks is given by
\begin{equation}
T(l_{\rm s}) =  \langle t (\tau+l_{\rm s}) - t(\tau) \rangle \simeq
\frac{1}{\tau_{\rm up}- \tau_{\rm dn}}
\sum_{\tau=\tau_{\rm dn}}^{\tau_{\rm up}} [t (\tau+l_{\rm s}) - t(\tau) ]\,.
\label{eq:Tl}
\end{equation}
Data for study of rotating strings and angular moments (see
sec.\ref{sec:rotstring}) were preformatted in a different way.  In
this case, the currency information has been projected onto the grid
of the regularly spaced 10~sec intervals.

\section{One dimensional maps}\label{sec:d1_map}

By applying standard methodologies of detrending one may suggest to
convert original series of the quotations of the  mean currency
exchange rate $p(\tau)$ onto a series of returns defined by
\begin{equation}
\frac{p(\tau+h)  -  p(\tau)}{p(\tau+h)}\,,
\end{equation}
where $h$ denotes a tick lag between currency quotes $p(\tau)$ and
$p(\tau+h)$, $\tau$ is the index of the quote. The mean
$p(\tau)=(p_{\rm ask}(\tau)+ p_{\rm bid}(\tau))/2$ is calculated
from $p_{\rm ask}(\tau)$ and $p_{\rm bid}(\tau)$.

In the spirit of the string theory it would be better to start with
the 1-end-point open string map
\begin{equation}
P^{(1)}(\tau,h) = \frac{p(\tau+h)  - p(\tau)}{p(\tau+ h)}\,,
\qquad h \in <0,l_{\rm s}>
\label{eq:string1}
\end{equation}
where superscript $(1)$ refers to the number of endpoints.

Later, we may use the notation $P\{p\}$ which emphasizes the
functional dependence upon the currency exchange rate $\{p\}$. It
should also be noted that the use of $P$ highlights the canonical
formal correspondence between the {\em rate of return} and the
internal {\em string momentum}.

\begin{figure}
 \begin{center}
 \includegraphics[height=16cm]{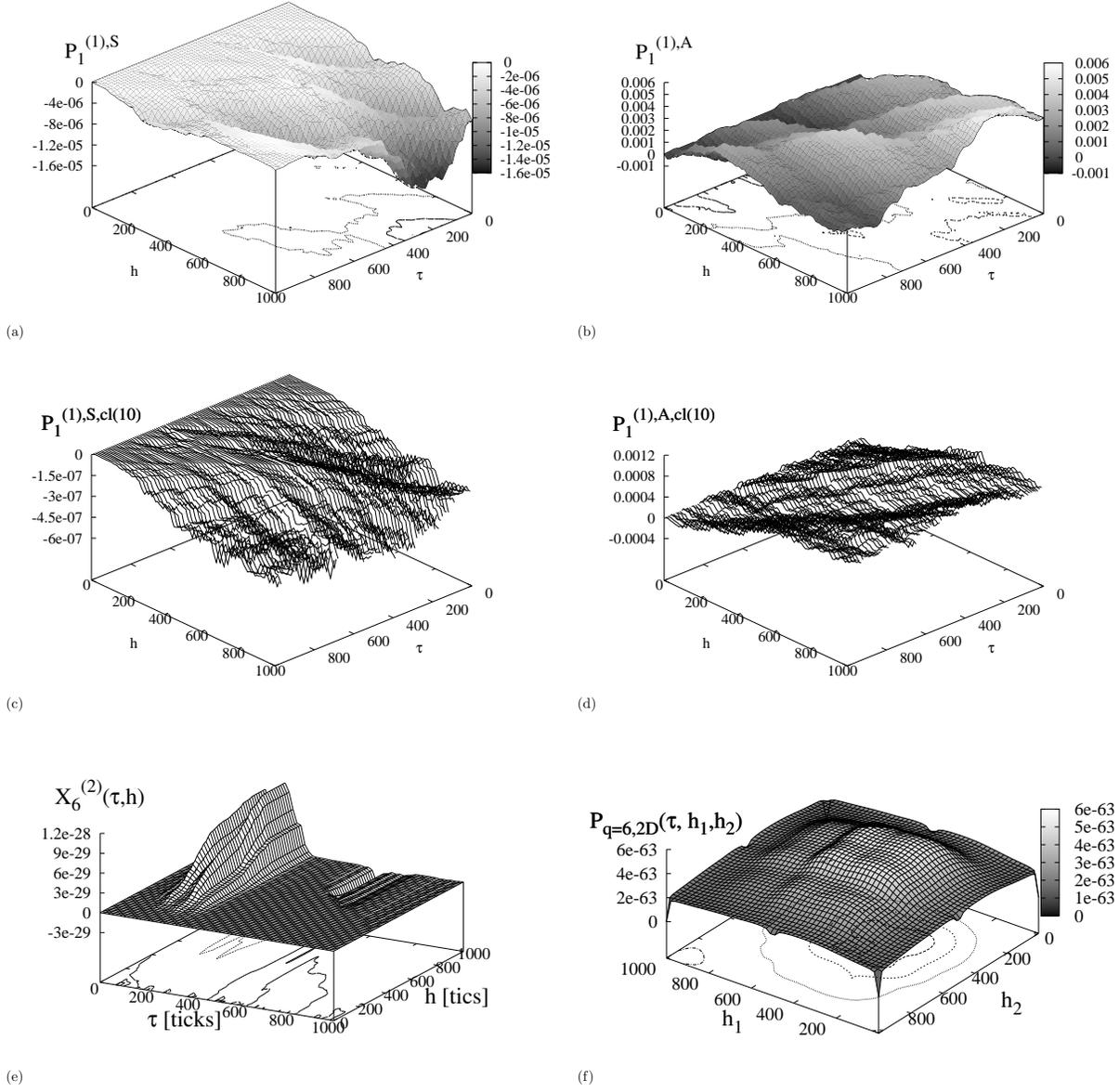}
 \caption{
 The illustrative examples of the currency data map for GBP/USD.
 The parts (a)-(d)
 constructed for date Fri, 31 Jul 2009 time interval
 15:06:37 - 15:43:09 GMT. Time evolution of symmetric
 ($P^{(1),\rm S}_{q=1}$) and anti-symmetric ($P^{(1),{\rm A}}$) component
 of the 1-end-point string of size $l_{\rm s}=1000$ calculated for $q=1$
 (by means of Eq.(\ref{eq:Psym})). In (c),(d) we see the
 same data mapped by means of the partially closed
 1-end-point string ($q=1$) for $N_{\rm m}=10$,  according to Eq.(\ref{eq:compact})).
 (e)~The calculation carried out for the 2-end-point string for $l_{\rm s}=1000$, $q=6$
 at some instant. We see that conjugate variable
 $X^{(2)}_{q=6}(\tau,h)$ satisfies the Neumann-type boundary conditions;
 (f)~The instantaneous 2D-Brane state
 (date Fri, 31 Jul 2009 15:11:47 GMT)
 is computed according to Eq.(\ref{eq:P2Dq}).}
 \label{fig:visual}
 \end{center}
 \end{figure}

\begin{figure}
 \begin{center}
 \includegraphics[width=16cm]{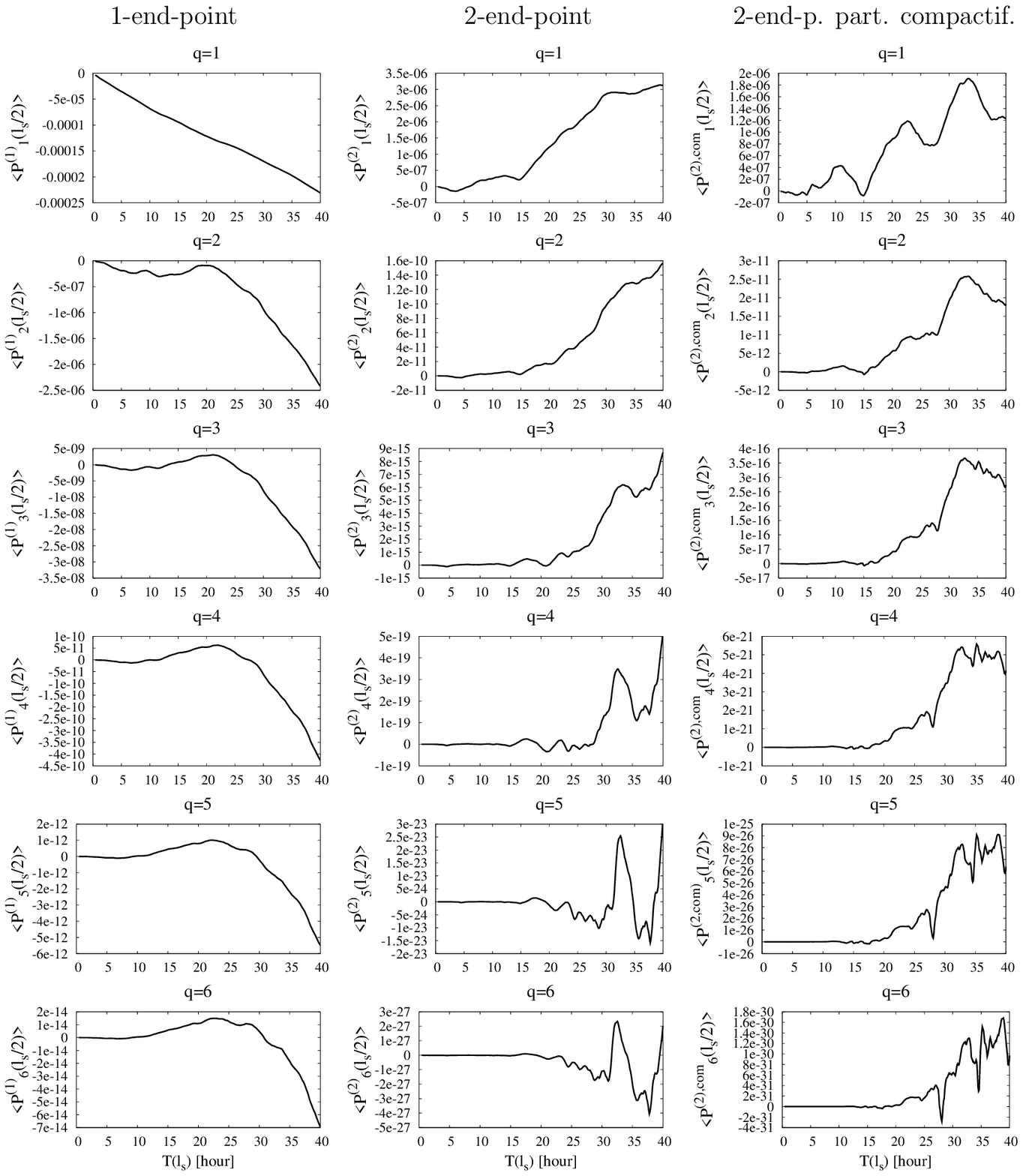}
 \caption{The variability in statistical characteristics
 caused by differences in topology and $q$. Calculated for the period Aug-Sep,
 GBP/USD currency. The model with $q=1$ has ability to reveal
 the currency long trend, on the other hand, the rare
 events are better visible for the 2-end-point string.
 The  effect of the partial compactification with
 $N_{\rm m}=4$ [see Eq.(\ref{eq:compact})] is demonstrated
 in the third column (again for the 2-end-point string).}
 \label{fig:1end_q1q6}
 \end{center}
 \end{figure}

Here the tick variable $h$ may be interpreted as a variable which
extends along the extra dimension limited by the string size $l_{\rm
s}$. A natural consequence of the transform, Eq.(\ref{eq:string1}),
is the fulfilment of the boundary condition
\begin{equation}
P^{(1)}(\tau,0)= 0\,,
\end{equation}
which holds for any tick coordinate $\tau$. Later on, we want to
highlight effects of the rare events. For this purpose, we introduce
a power-law q-deformed model
\begin{equation}
P^{(1)}_q(\tau,h) =
f_q \left( \frac{p(\tau+h)- p(\tau)}{p(\tau+ h)}\right)\,, \qquad h \in <0,l_{\rm s}>
\label{eq:stringq}
\end{equation}
by means of the function
\begin{equation}
f_q(x)=  \mbox{sign}(x)\,|x|^q\,,\qquad q>0\,.
\end{equation}
The 1-end-point string has defined the origin, but it reflects the
linear trend in $p(.)$ at the scale $l_{\rm s}$. Therefore, the
1-end-point string map $P^{(1)}_q(.)$ may be understood as a
q-deformed generalization of the {\em currency returns}. The
illustration of the 1-end-point model is given in
Fig.(\ref{fig:visual}). The corresponding statistical
characteristics displayed in Fig.(\ref{fig:1end_q1q6}) have been
obtained on the basis of a statistical analysis discussed in
section~\ref{sec:Data}.

Clearly, the situation with a long-term trend is partially corrected
by fixing $P^{(2)}_q(\tau,h)$ at $h=l_{\rm s}$. The open string with
two end points is introduced via the nonlinear map which combines
information about trends of $p$ at two sequential segments
\begin{equation}
P^{(2)}_q(\tau,h)  =
f_q\left( \left(\frac{ p(\tau+h)  -  p(\tau) }{p(\tau + h)}  \right)
\left( \frac{ p(\tau+l_{\rm s}) -  p(\tau+h) }{p(\tau +l_{\rm s})} \right)
\right) \,,\qquad  h \in  <0,l_{\rm s}>\,.
\label{eq:Pq2p}
\end{equation}
The map is suggested to include boundary conditions of {\em Dirichlet type}
\begin{equation}
P^{(2)}_q(\tau,0)= P_q(\tau,l_{\rm s})=0\,,\qquad
\mbox{at all ticks}\,\,\, \tau\,.
\label{eq:Dir}
\end{equation}
In particular, the sign of $P^{(2)}_q(\tau,h)$ comprises
information about the behavior differences
of $p(.)$ at three quotes $(\tau,\, \tau+h,\, \tau+l_{\rm s})$.
The $P^{(2)}_q(\tau,h)<0$ occurs for trends of the different
sign, whereas $P^{(2)}_q(\tau,h)>0$ indicates the match
of the signs.

In addition to the variable $P^{(2)}_q(\tau,h)$ we introduced the
conjugate variable $X^{(2)}_q(\tau)$ via the recurrent summation
\begin{equation}
X^{(2)}_q(\tau,h+1) = X^{(2)}_q(\tau,h)   +
P^{(2)}_q(\tau, h-1)\,[\,t(\tau+h)- t(\tau + h - 1)\,]\,
\label{eq:Xq2}
\end{equation}
(here $t(.)$ stands for a time-stamp corresponding to the quotation
index $\tau$ in the argument). The above discrete form is suggested
on the basis of the time-continuous Newton second law of motion
$\dot{X}^{(2)}_q(t,h)=P^{(2)}_q(t,h)$ (written here for a unit
mass). The form is equivalent to the imposing of the quadratic
kinetic energy term $\frac{1}{2} (P^{(2)}_q)^2 $. Thus, the
Hamiltonian picture \cite{Green1987} can be reconstructed in the
following way;
\begin{equation}
{\cal H}=  \frac{1}{2}
\sum_{h=0}^{l_{\rm s}} \left[ (\,P^{(2)}_q(\tau,h)\,)^2 -
\left[\phi_{\rm ext}(\tau,h+1)  -  \phi_{\rm ext}(\tau,h)\right]\, X^{(2)}_q(\tau,h)\, \right]\,,
\end{equation}
where $\phi_{\rm ext}(\tau,h)$ is the external field term which
depends on the transform of the currency rate~[see e.g.
Eq.(\ref{eq:Pq2p})]. We pass from the continuum to discrete theory
by means of the functional form
\begin{eqnarray}
\dot{P}^{(2)}_q &=& - \frac{\delta {\cal H}}{\delta
X^{(2)}_q(h)} =  {\phi}_{\rm ext}(\tau,h+1) - \phi_{\rm ext}(\tau,h)
=  P^{(2)}_q(\tau,h+1)- P^{(2)}_q(\tau,h)\,,
\end{eqnarray}
where $P^{(2)}_q(\tau,h)$ can be calibrated equal to $\phi_{\rm
ext}(\tau,h)$.

The discrete conjugate variable meets the {\em Neumann type}
boundary conditions
\begin{equation}
X^{(2)}_q(\tau,0) = X^{(2)}_q(\tau,1)\,,
\qquad X^{(2)}_q(\tau,l_{\rm s}-1) = X^{(2)}_q(\tau,l_{\rm s})\,,
\end{equation}
which is illustrated in Fig.(\ref{fig:visual})(d).

A more systematic way to obtain the 2-end-point string map
represents the method of undetermined coefficients. The numerator of
$q=1$ can be chosen in the functional polynomial form of degree 2
with coefficients $\beta_0, \ldots, \beta_5$ as follows:
\begin{eqnarray}
P^{(2)}_{q=1,\rm Num}(\tau,h)&=&
\beta_0 p^2(\tau+h) +
\beta_1 p^2(\tau)+\beta_2 p^2(\tau+l_{\rm s})
\\
&+&
\beta_3 p(\tau)p(\tau+h) + \beta_4 p(\tau) p(\tau+l_{\rm s}) +
\beta_5 p(\tau+h)p(\tau+l_{\rm s})\,.
\end{eqnarray}
Again, the Dirichlet conditions $P^{(2)}_{{q=1},{\rm Num}}(\tau,0)=
P_{{q=1},{\rm Num}}(\tau, l_{\rm s})=0$ yield $P^{(2)}_{{q=1},{\rm
Num}}= \beta_0 ( p(\tau) - p(\tau+h) ) ( p(\tau+l_{\rm s})-
p(\tau+h))$ with arbitrary $\beta_0$. The overlooked denominator
part of fraction $P_{q=1}^{(2)}$ then servers as a normalization
factor.

Another interesting issue is the generalizing 1-end-point string to
include the effect of many length scales
\begin{equation}
P^{(N_{l_{\rm s}})}_q(\tau,h;\{l\}) = \prod_{i=1}^{N_{l_{\rm s}}}
f_q\left(\,\frac{p(\tau+l_i)  -  p(\tau+h)}{p(\tau+h)}\right)\,,
\end{equation}
which relies on the sequence $\{l\} \equiv$ $\{ \, l_i$, $i=1,
\ldots, N_{l_{\rm s}}\}$, including the end points
($\min_{i=1,\ldots, N_{l_{\rm s}}} l_i$ and $\max_{i=1,\ldots,
N_{l_{\rm s}}} l_i$) as well as the $N_{l_{\rm s}}-2$ interior node
points that divide the string map into the sequence of unfixed
segments of the non-uniform length (in general).

\section{Symmetry with respect to $p(.)\rightarrow 1/p(.)$ transform}\label{sec:sym}

The currency pairs can be separated into {\em direct} and {\em
indirect} type. In a direct quote the {\em domestic} currency is the
base currency, while the {\em foreign} currency is the quote
currency. An {\em indirect} quote is just the opposite. Therefore,
it would be interesting to take this symmetry into account. Hence,
one can say that this two-fold division of the market network admits
{\em duality symmetry}. Duality symmetries are some of the most
interesting symmetries in physics. The term {\em duality} is used to
refer to the relationship between two systems that have different
descriptions but identical physics (identical trading operations).

Let us analyze the 1-end-point elementary string map when the
currency changes from direct to indirect. The change can be
formalized by means of the transformation
\begin{equation}
\hat{\mathcal T}_{\rm id}: \,
P\{p(.)\}  \rightarrow  \overline{P}\{p(.)\} \equiv P\{1/p(.)\}\,,
\label{TPp}
\end{equation}
For the 1-end-point map model of the string, Eq.(\ref{eq:stringq}),
we obtained
\begin{equation}
\hat{\mathcal T}_{\rm id} P^{(1)}_q(\tau,h) =
\overline{P}^{(1)}_q(\tau,h)
=  f_q \left( \frac{p(\tau)  -   p(\tau+h)}{p(\tau)} \right)\,.
\label{eq:stringqT}
\end{equation}
Let us consider two-member space of maps $V^{(1)}_P= \{P^{(1)}_q$,
$\overline{P}^{(1)}_q\}$. Important, we see that $\hat{\mathcal
T}_{\rm id}$ preserves the Dirichlet boundary conditions, in
addition, the identity operator $\hat{\mathcal T}^2_{\rm id}$ leaves
the elements of $V^{(1)}_P$ unchanged. The space $V^{(1)}_P$ is
closed under the left action of $\hat{\mathcal T}_{\rm id}$. These
ideas are straightforward transferable to the 2-end-point string
points.

Now we omit the notation details and proceed according to
Eq.(\ref{TPp}). The map $P(.)$ is decomposable into a sum of
symmetric and antisymmetric parts
\begin{equation}
P^{\rm S} =   \frac{1}{2} ( P + \overline{P}  )\,, \qquad P^{\rm A}
=   \frac{1}{2} ( P - \overline{P}  ), \label{eq:Psym}
\end{equation}
respectively. Due to of normalization by $1/2$, we get the
projection properties
\begin{equation}
\hat{\mathcal T}_{\rm id} P^{\rm S} =    P^{\rm S}\,,
\qquad
\hat{\mathcal T}_{\rm id} P^{\rm A} =  - P^{\rm A}\,.
\end{equation}
To be more concrete, we choose $q=1$ and obtain
\begin{equation}
P_{q=1}^{(1),{\rm S}}
=  1 - \frac{1}{2}
\left[ \frac{p(\tau)}{p(\tau+h)}  +  \frac{p(\tau+h)}{p(\tau)} \right]\,,
\qquad
P_{q=1}^{(1),{\rm A}} =
\frac{1}{2} \left[ \frac{p(\tau)}{p(\tau+h)}
-\frac{p(\tau+h)}{p(\tau)}\right]\,.
\label{eq:Pq1pS}
\end{equation}
and
\begin{eqnarray}
P_1^{(2),{\rm A}}  &=&
 \frac{1}{2}\left[
 \frac{p(\tau)}{p(\tau+l_{\rm s})} -
 \frac{p(\tau+l_{\rm s})}{p(\tau)} +
 \frac{p(\tau+h)}{p(\tau)}  -
 \frac{p(\tau)}{p(\tau+h)}  +
 \frac{p(\tau+l_{\rm s})}{p(\tau+h)} - \frac{p(\tau+h)}{p(\tau+l_{\rm s})}
 \right]\,,
\label{eq:Pq1pS2}
\\
P_1^{(2),{\rm S}}  &=&  1 + \frac{1}{2}
\left[ \frac{p(\tau+l_{\rm s})}{p(\tau)}
+ \frac{p(\tau)}{p(\tau+l_{\rm s})}
- \frac{p(\tau)}{p(\tau+h)} - \frac{p(\tau+h)}{p(\tau)}
- \frac{p(\tau+h)}{p(\tau+l_{\rm s})}
- \frac{p(\tau+l_{\rm s})}{p(\tau+h)} \right]\,.
\nonumber
\end{eqnarray}
We see that the $P_{q=1}^{(1),{\rm S}}$ and $P_{q=1}^{(2),{\rm S}}$
maps acquire formal signs of the systems with $T-${\em dual
symmetry}~\cite{Zwie2009}. When the world described by the closed
string of the radius $R$ is indistinguishable from the world of the
radius $\propto 1/R$ for any $R$, the symmetry manifests itself by
$(R \pm \mbox{const.}/R)$ terms of the mass squared operator. The
correspondence with our model becomes apparent one assumes that $R$
corresponds to the ratio $p(\tau)/p(\tau+h)$ in Eq.(\ref{eq:Pq1pS}).
However, we must also refer a reader to an apparently serious
difference that in our model we do not consider for the moment the
compact dimension.

\subsection{${\mathcal T}_{\rm id}$ transform under the conditions of bid-ask spreads}

Simply, the generalization can also be made with allowing for
currency variables which appear as a consequence of the transaction
costs \cite{Wagner1993}. The occurrence of ask-bid spread
complicates the analysis in several ways. Instead of one price for
each currency, the task requires the availability to two prices. The
impact of ask-bid spread on the time-series properties has been
studied within the elementary model~\cite{Roll1984}.

Thus, for the purpose of a thorough and more realistic analysis of
the market information, it seems straightforward to introduce
generalized transform
\begin{equation}
\hat{\mathcal T}^{\rm ab}_{\rm id}
P
\{\,p_{\rm ask}(.),p_{\rm bid}(.)\,\}
= \overline{P}\{ \,
1/p_{\rm bid}(.),
1/p_{\rm ask}(.)\,\}\,,
\end{equation}
which converts to Eq.(\ref{eq:stringqT}) in the limit of vanishing spread.

\section{Mapping to the model of 2D Brane}\label{sec:Brane}

Clearly, there is a possibility to go beyond a string model towards
more complex maps including alternative spread-adjusted currency
returns. Formally, the generalized mapping onto the 2D brane with
the $(h_1, h_2) \in <0,l_{\rm s}> \times <0,l_{\rm s}>$ coordinates
which vary along two extra dimensions could be proposed in the
following form:
\begin{eqnarray}
P_{{\rm 2D},q}(\tau, h_1, h_2)  &=&  f_q \left(
\left(\frac{
p_{\rm ask}(\tau + h_1) -  p_{\rm ask}(\tau) }{
p_{\rm ask}(\tau + h_1)}  \right) \,
\left( \frac{ p_{\rm ask}(\tau+l_{\rm s}) -
p_{\rm ask}(\tau+h_1) }{p_{\rm ask}(\tau +l_{\rm s})} \right)
\right.
\label{eq:P2Dq}
\\
&\times &
\,\,\,\,\,\,\,\,\,
\left.
\left(\frac{
p_{\rm bid}(\tau) -
p_{\rm bid}(\tau+h_2) }{p_{\rm bid}(\tau )}  \right) \,
\left( \frac{ p_{\rm bid}(\tau+h_2) -
p_{\rm bid}(\tau+l_{\rm s}) }{p_{\rm bid}(\tau+h_2)} \right)
\right) \,.
\nonumber
\end{eqnarray}
The map constituted by the combination of "bid" and "ask" quotes is
constructed to satisfy the Dirichlet boundary conditions
\begin{eqnarray}
P_{{\rm 2D},q}(\tau, h_1, 0)   =   P_{{\rm 2D},q}(\tau, h_1, l_{\rm s})   =
P_{{\rm 2D},q}(\tau, 0,  h_2)  =   P_{{\rm 2D},q}(\tau, l_{\rm s}, h_2) \,.
\end{eqnarray}
In addition, the above construction, Eq.(\ref{eq:P2Dq}), has been
chosen as an explicit example, where the action of $\hat{\mathcal
T}^{\rm ab}_{\rm id}$ becomes equivalent to the permutation of
coordinates
\begin{eqnarray}
\hat{\mathcal T}^{\rm ab}_{\rm id} P_{{\rm 2D},q}(\tau, h_1, h_2) =  P_{{\rm 2D},q}(\tau,h_2,h_1)\,.
\end{eqnarray}
Thus, the symmetry with respect to interchange of extra dimensions
$h_1, h_2$ can be achieved through $P_{{\rm 2D},q} + \hat{\mathcal
T}^{\rm ab}_{\rm id} P_{{\rm 2D},q}$.  In a straightforward
analogous manner one can get an antisymmetric combination $P_{{\rm
2D},q} -  {\hat{\mathcal T}}^{\rm ab}_{\rm id} P_{{\rm 2D},q}$. For
a certain instant of time we proposed illustration which is depicted
in Fig.(\ref{fig:visual})(b).

At the end of this subsection, we consider the next even simple
example, where mixed boundary conditions take place. Now let the
2-end-point string be allowed to pass to the 1-end-point string by
means of the homotopy $P_{q_1,q_2}^{(1,2)}(\tau, h, \eta)= (1-\eta)
P^{(1)}_{q_1}(\tau,h) +  \eta P^{(2)}_{q_2}(\tau,h)$ driven by the
parameter $\eta$ which varies from $0$ to $1$.  In fact, this model
can be seen as a variant of the 2D brane with extra dimensions $h$
and $\eta$.

\subsection{Partial compactification} \label{sec:partial}

In the frame of the string theory, the compactification attempts to
ensure compatibility of the universe based on the four observable
dimensions with twenty-six dimensions found in the theoretical model
systems.  From the standpoint of the problems considered here, the
compactification may be viewed as an act of the information
reduction of the original signal data, which makes the transformed
signal periodic. Of course, it is not very favorable to close
strings by the complete periodization of real input signals. Partial
closure would be more interesting. This uses pre-mapping
\begin{equation}
\tilde{p}(\tau) = \frac{1}{N_{\rm m}}
\sum_{m=0}^{N_{\rm m}-1} p(\tau +l_{\rm s} m)\,,
\label{eq:compact}
\end{equation}
where the input of any open string (see e.g. Eq.(\ref{eq:string1}), Eq.(\ref{eq:Pq2p}))
is made up partially compact.

Thus, data from the interval $<\tau, \tau+l_{\rm s} ( N_{\rm m}-1 )
>$ are being pressed to occupy "little space" $h\in <0, l_{\rm s}>$.
We see that as $N_{\rm m}$ increases, the deviations of $\tilde{p}$
from the periodic signal become less pronounced. The idea is
illustrated in Fig.(\ref{fig:visual})(c),(d). We see that the states
are losing their original form~(a),(b) are starting to create
ripples.

For example, one might consider the construction of the
$(\tilde{D}+1)$-brane
\begin{equation}
f_q \, \left( \frac{p(\tau + h_0)- p(\tau)}{p(\tau+h_0)}
\right)
\,
\prod_{j=1}^{\tilde{D}}
f_q\left( \frac{\tilde{p}^{(\pm)}_j(\tau + h_j) - \tilde{p}^{(\pm)}_j(\tau)}{ {\tilde p}^{(\pm)}_j(\tau+h_j) }
\right)
\end{equation}
maintained by combining $(\tilde{D}+1)$ 1-end-point strings,
where partial compactification in $\tilde{D}$ extra dimensions
is supposed.  Of course, the construction introduces auxiliary variables
$\tilde{p}^{(\pm)}_j(\tau) = \sum_{m=0}^{N_{\rm m},j-1} p(\tau \pm m\,l_{{\rm s},j})$.

\section{Statistical investigation of 2-end-point strings}\label{sec:stat}

\subsection{The midpoint information about string}\label{sec:midpoint}

In our present work, the strings and branes represent targets of
physics-motivated maps which convert an originally dynamic range of
currency data into the static frame. Of course, the data shaped by
the string map have to be studied by the statistical methods.
However, the question remains open about the selection of the most
promising types of maps from the point of view of interpretation of
their statistical response.

\begin{figure}
 \begin{center}
 \includegraphics[width=17cm]{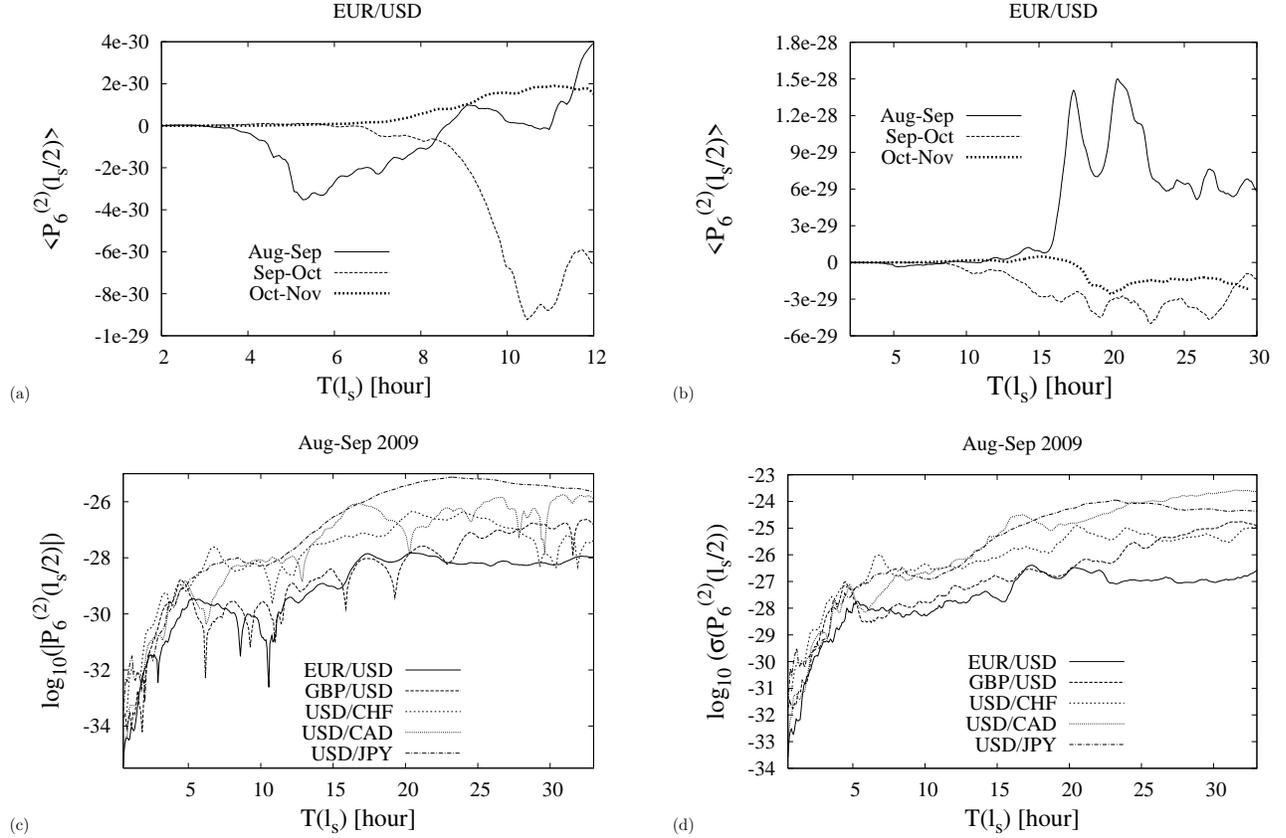}
 \caption{ The illustrative calculations carried out for EUR/USD currency.
 Figure shows the parts~(a),(b) which include a view of two different epochs
 (and their different details).  We see the variability of the mean statistical
 characteristics of the 2-end-point open string.
 The part (b)~turns in sign,
 but remarkable exceptional scales corresponding to the local maxima and minima remain the same.
 The string length is expressed in real-time units
 calculated by means of Eq.(\ref{eq:Tl}). In part (c)~we
 present anomalies - peaks roughly common for
 different currencies. These picture of anomalies
 are supplemented by dispersions of
 $P^{(2)}_q(l_{\rm s}/2)$~(d).}
 \label{fig:reson}
 \end{center}
 \end{figure}

Many of the preliminary numerical experiments we performed
indicating that the 2-end-point strings with a sufficiently high $q$
(in this work we focus on $q=6$, but other unexplored values may
also be of special interest) yield interesting statistical
information including focus on rare events. Unfortunately, there is
difficult or impossible to be exhaustive in this aspect.
Figure(\ref{fig:reson}) shows how $<P^{(2)}_6(\tau,h)>$ and the
corresponding dispersion $\sigma_{P_6}$ change with a string length.

 \subsection{The analysis of $P_q(l_{\rm s}/2)$ distributions}\label{sec:distrib}

\begin{figure}
 \begin{center}
 \includegraphics[width=16cm]{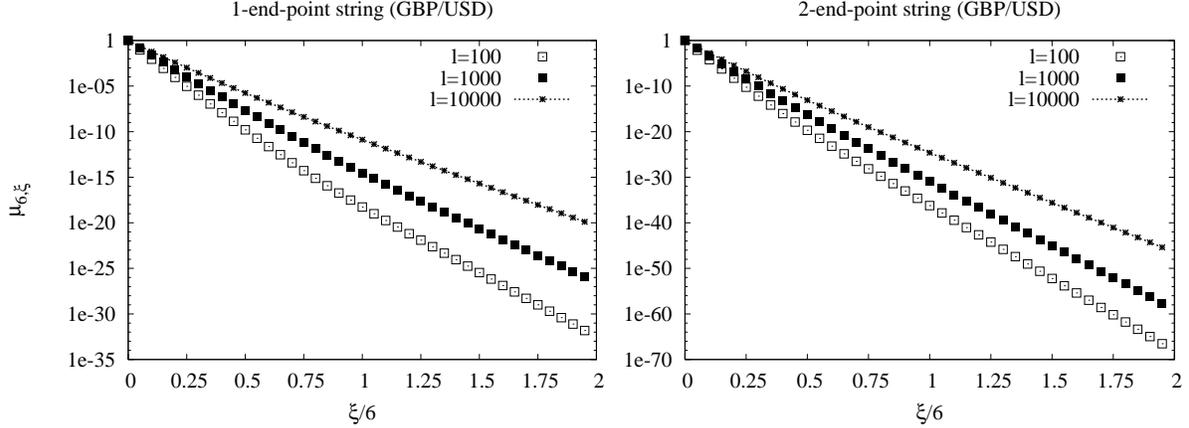}
 \caption{The mid-point fluctuations characterized by the statistical
 moments defined by Eq.(\ref{eq:momxi}). The calculations are carried out for
 GBP/USD currency rate, for {\rm Aug-Sep} period, for different kinds of
 strings for several lengths. We see
 that fluctuations become more significant as the string size
 increases. In addition, one may observe the 2-end-point string
 to be more suppressive to the fluctuations.}
 \label{fig:endpoint}
 \end{center}
 \end{figure}

 The complex trade fluctuation data can be characterized
 by their respective statistical moments.
 In the case of the string map the moments
 of the $\xi$th order can be naturally considered
 at the half length
 \begin{equation}
 \mu_{q,\xi} = \langle  \,|P^{(N_l)}_q(\tau, l_{\rm s}/2)|^{\xi/q} \rangle \,.
 \label{eq:momxi}
 \end{equation}
 The comparison of the results obtained for the 1-end
 point and 2-end point strings is depicted in
 Fig.\ref{fig:endpoint}.  The remarkable difference in the amplitudes
 is caused by the manner of anchoring.
 The moments of longer strings are trivially larger.

 \subsection{Volatility vs. string amplitude}\label{sec:volat}

 \begin{figure}
 \begin{center}
 \includegraphics[width=16cm]{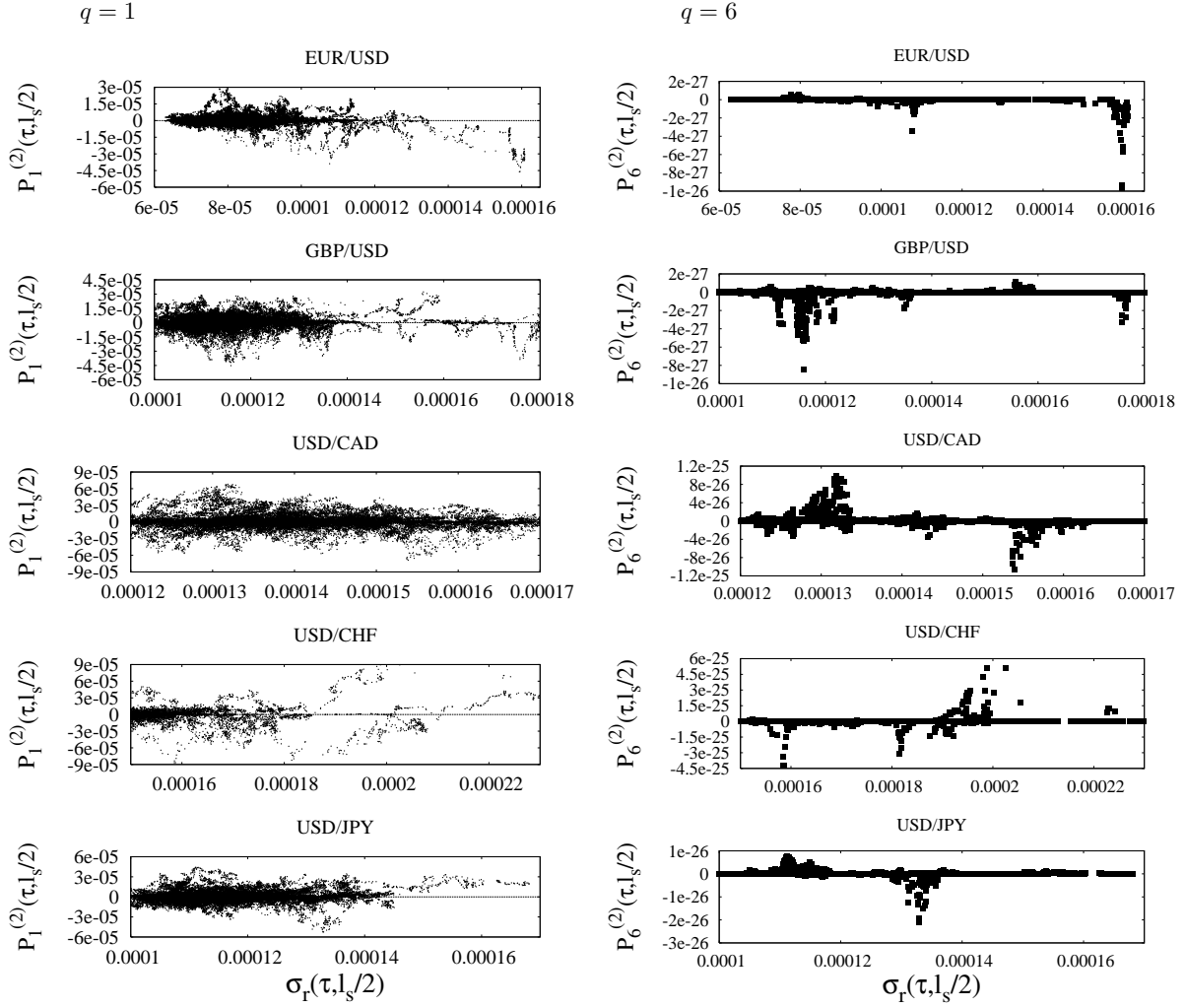}
 \caption{
 The scatterplot showing the relationship between the volatility
 $\sigma_r(\tau,l_{\rm s}/2)$ and the string amplitudes $P_1^{(2)}(l_{\rm s}/2)$ ($q=1$)
 and $P_6^{(2)}(l_{\rm s}/2)$ ($q=6$), respectively. The separating
 effect at high $q$ is visible. The plot indicates conservation
 or brake of the price trend $P_6^{(2)}(l_{\rm s}/2)$
 over the tick time $<\tau, \tau+l_{\rm s}>$.
 We see that the trend becomes coupled
 with the occurrence of specific isolated
 values of the volatility calculated for $l_{\rm s}=10000$; period Aug-Sep.}
 \label{fig:volat}
 \end{center}
 \end{figure}

The volatility as described here refers to the standard deviation of
currency returns of a financial instrument within a specific time
horizon described by the length $l_{\rm s}/2$. The return volatility
at the time scale $l_{\rm s}/2$ is defined by $\sigma_r(l_{\rm
s}/2)= \sqrt{r_2(l_{\rm s}/2)-r_1^2(l_{\rm s}/2)}$ using $r_m(l_{\rm
s})=\sum_{h=1}^{l_{\rm s}/2} [(p(\tau+h)-p(\tau+h-1))/p(\tau+h)]^m$
for $m=1, 2$. In Fig.(\ref{fig:volat}), the {\em rate of return
volatility} computed at the scale $L=l_{\rm s}/2$ demonstrates the
linkage to the changes in the price trend represented by
$P_6^{(2)}(l_{\rm s}/2)$. Since the trend changes do not follow
Gaussian distributions, we have used high $q$  to analyze the impact
of rare events. In Fig.(\ref{fig:volat}), we show the identification
of the semi-discrete levels of volatility by $q=6$, while setting
$q=1$ does not uncover common attributes.

\section{Intra-string statistical picture}\label{sec:intrastring}

 \begin{figure}
 \begin{center}
 \includegraphics[width=16cm]{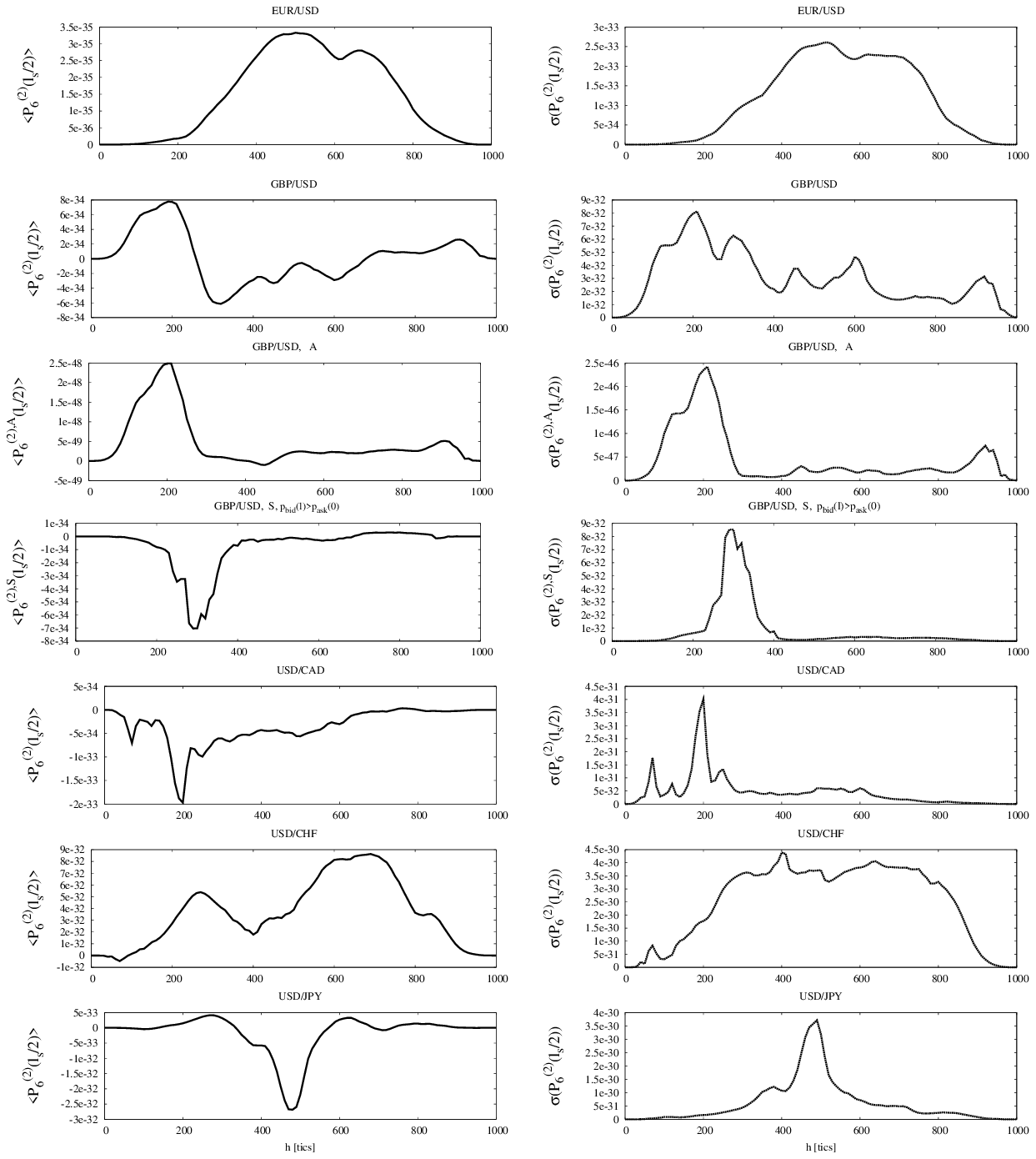}
 \caption{The intra-string statistics for five currency pairs during the Aug-Sep
 period; the 2-end-point string is studied for $q=6$, $l_{\rm s}=1000$.
 The mean $<P^{(2)}_6(l_{\rm s}/2)>$
 (l.h.s. column) and dispersion (r.h.s. column) of  $P^{(2)}_6(\tau,h)$.
 The specific scope brings analysis of the effects of symmetry and spread
 for statistics collected within constraint that GBP/USD base currency is trending up
 ($p_{\rm bid}(l_{\rm s}) \geq p_{\rm ask}(0)$).
 The amplitude of $<P_6^{(2),{\rm A}}(l_{\rm s}/2)>$
 is very small (fluctuations are suppressed) as compared to the average of
 the symmetric part.  On the other hand, $<P_6^{(2),{\rm S}}(l_{\rm s}/2)>$
 is not taken into plot since it is indistinguishable from
 $<P_6^{(2)}(l_{\rm s}/2)>$.}
 \label{fig:P_disper}
 \end{center}
 \end{figure}

The idea of the string related maps proposed here is the
transformation of the original point object such as selected single
price into a system of a prices from its admissible neighborhood.
This changeover from a local to a non-local description directly
extends econometrics belief that future prices are deducible from
the price history of a given period. The results of the
investigation of the intra-string statistics for a selected scale
$l_{\rm s}$ (period) are presented in Fig.(\ref{fig:P_disper}).

\subsection{Time invariant strings and elementary statistics}\label{sec:exp}

To understand better the outputs of our numeric attempts, let us
suppose to deal with short-time evolution of the currency $\ln
p(\tau)  =  \ln p_0 + b \tau$ characterized by the
linear-logarithmic parameter $b$.  Surprisingly, after the
substitution into the string, Eq.(\ref{eq:Pq2p}), the expressions
collapse to the invariant (independent of $\tau$) form
\begin{eqnarray}
P^{(1)}_1(\tau,h)&=& 1 - \exp(-h b)\,,
\label{eq:P12inv}
\\
P^{(2)}_1(\tau,h)&=& 1 - \exp[ (h-l_{\rm s}) b] -
\exp(-h b) + \exp(-l_{\rm s} b)\,.
\nonumber
\end{eqnarray}
It is quite interesting to look at the lowest order terms of Taylor
series of this result around $b=0$. Ignoring the terms of order
$b^4$ or higher gives $P_1^{(1)} =  b h - \frac{1}{2} b^2 h^2 +
{\cal O}(b^3)$, $P_1^{(2)} =  b^2 h ( l_{\rm s}- h )  +  {\cal
O}(b^3)$. It demonstrates that the model of the 1-end-point string
is more sensitive to the sign of $b$ variations.

The elementary qualitative statistical model of the string can be
obtained by taking the unexamined assumption that $b$ fluctuates
with the Gaussian probability density $(2\pi\sigma_{\rm b}^2)^{-1}
\exp[ -b^2/(2\sigma_{\rm b}^2)]$. The averaging of
Eq.(\ref{eq:P12inv}) with this weight yields
\begin{eqnarray}
<P^{(1)}_1>_{\tiny\mbox{Gauss}}(h) &=&1 -  \exp\left(-\frac{h^2\sigma_{\rm b}^2}{2}\right)\,,
\label{eq:sigm1}
\\
<P^{(2)}_1>_{\tiny\mbox{Gauss}}(h) &=&1  -  \exp\left(-\frac{h^2 \sigma_{\rm b}^2}{2}\right)-
\exp\left(-\frac{( h - l_{\rm s})^2  \sigma_{\rm b}^2}{2} \right) + \exp\left(- \frac{l_{\rm s}^2 \sigma_{\rm b}^2}{2} \right)\,.
\noindent
\end{eqnarray}
Equation(\ref{eq:sigm1}) predicts an increase in the $h$ dependence
as a consequence of the symmetric fluctuations in $b$. This finding
does not agree with the result presented in
Fig.(\ref{fig:1end_q1q6}), where the mean string output is biased by
the mean trend.

\subsection{Mapping of the periodic input signal} \label{sec:per}

Simultaneously with giving numerical results obtained for
statistical averages of data, it is instructive to briefly examine
the string map of the signal of periodic form described by some
elementary function. The input signal $p(\tau) =  a_1 + a_2
\cos(\omega \tau)$ can be suitable for this purpose. Subsequently,
the analytic calculation for the 2-end-point map can be carried out
perturbatively under the requirement $a_2 \ll a_1$.  The common form
of average unifying formulas obtained for different integer $q$
values can be written as
\begin{equation}
<P^{(2),(\rm S)}_q(h)>_{\rm cos}  =
\left[\cos(h\omega) +
\cos((h-l_{\rm s})\omega) -
\cos(l_{\rm s}\omega)-1 \right]^q
\sum_{j=0}
c_{q,j}\left(\frac{a_2}{a_1}\right)^{2(q + j)}\,,
\end{equation}
where $c_{q,j}$ are the numerical coefficients which are not
critical for further reasoning.  The intuitive idea that
$P^{(2),(\rm A)}_q(\tau,h)$ discriminates fluctuations, which stems
from the comparison of the symmetric and antisymmetric averages (see
Fig.(\ref{fig:P_disper})), is partially justified by the result $<
P^{(2),(\rm A)}_q(h)>_{\rm cos} =0$.

Interestingly, the calculation highlights the idea of the presence
of the resonant lengths $l_{\rm s}(n) = 2\pi n/\omega$ as
$n=1,2,\ldots$. This basic result motivated us to introduce the
2-end-point string model, that has potential to identify
characteristic dynamic scales represented here by $1/\omega$. It is
important that anomalous aspect is absent in the statistical
characteristics of the one-point strings.

\subsection{String map in the representation of internal Fourier modes}\label{sec:DFT}

\begin{figure}
  \begin{center}
  \includegraphics[width=16cm]{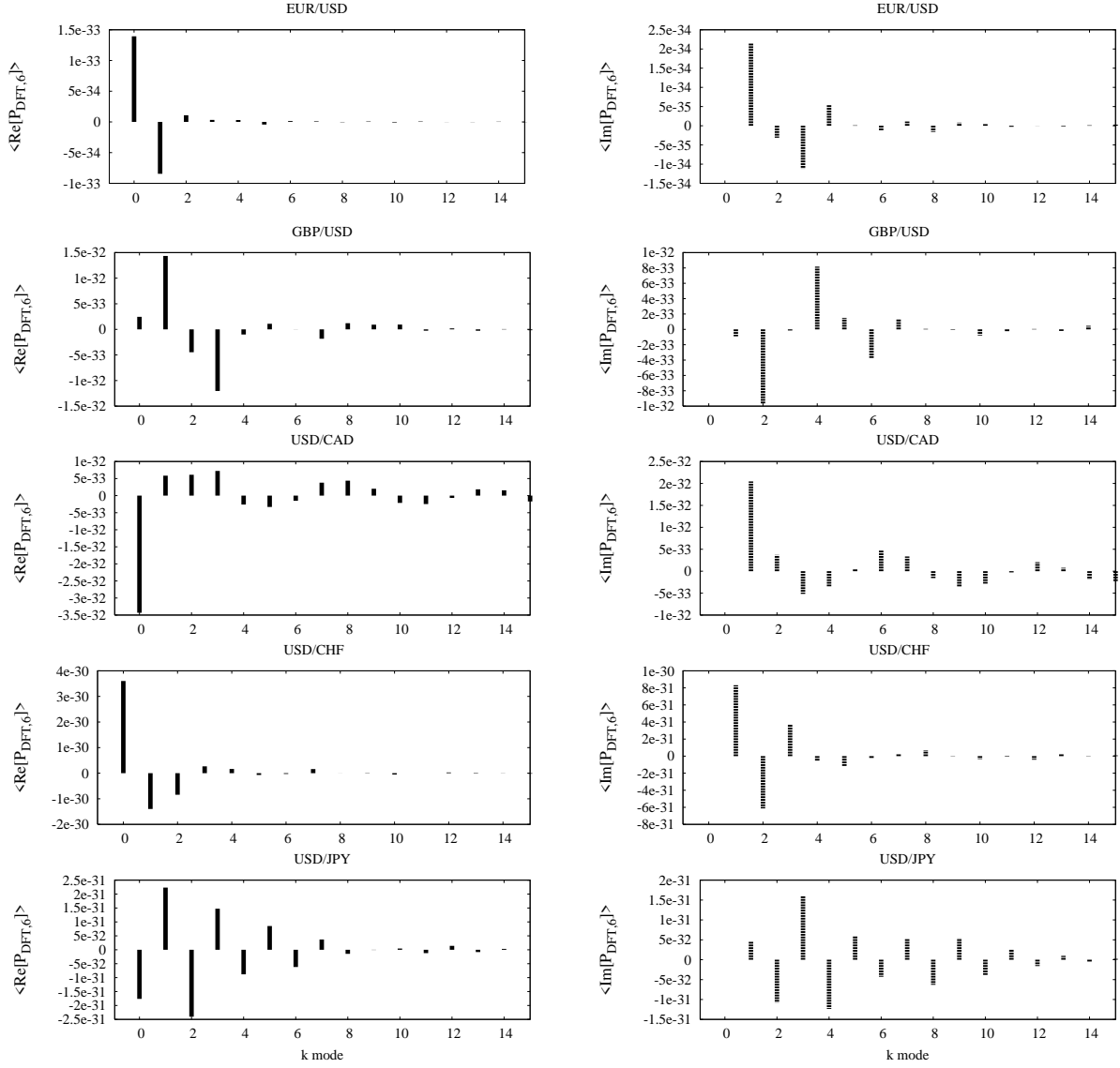}
  \caption{The currency-specific slow ($k\leq 14$) intra-string averaged
  Fourier modes (their real and imaginary part separately)
  obtained by the temporal averaging over the Aug-Sep period.
  The mean real and imaginary parts calculated for five pairs,
  $q=6$ and string length $l_{\rm s}=1000$.}
 \label{fig:DFT}
 \end{center}
 \end{figure}

At each tick $\tau$ the string may be represented by the sequence
$P^{(2)}_q(\tau,h)$, $h=0, 1, \ldots, l_{\rm s}-1$ which can be
transformed by means of the discrete Fourier transform
\begin{equation}
P_{{\rm DFT},q}(k,\tau) =
\sum_{h=0}^{l_{\rm s}} P^{(2)}_q(\tau, h) \exp\left( -
\frac{2\pi i k h}{l_{\rm s}+1}
\,\right)\,,
\quad  k= 0, 1, \ldots, l_{\rm s}\,.
\end{equation}
Having done this, one can introduce the inverse transform
\begin{equation}
P_{{\rm IDFT},q}(h,\tau)  =
\sum_{k=0}^{l_{\rm s}} P_{{\rm DFT},q}(\tau,k) \exp
\left( \frac{2 \pi i k h}{l_{\rm s}+1} \right)\,,
\end{equation}
which can be understood as a periodic extension of the input
$P^{(2)}_q(\tau,h)$ with a period of the $(l_{\rm s}+1)$ ticks. Then
$P_{{\rm IDFT},q}(h,\tau)$ can be viewed as a portion of the
original signal which curls up along the {\em compact dimension} of
the {\em closed string}. Thus, the integer $h/(l_{\rm s}+1)$ has the
meaning of a winding number of $P_{{\rm IDFT},q}(h,\tau)$.  The
statistical mean values of $P_{{\rm DFT},q}(k,\tau)$  calculated for
different currencies are depicted in Fig.(\ref{fig:DFT}). The
Fourier transform of the inherent  string structure may serve to
identify distinguishing features of currencies at selected time
scale.

\section{String polarized by the external field}\label{sec:polar}

In this section, we modify the two-point-string model,
Eq.(\ref{eq:Pq2p}), in order to account for the transaction costs. A
natural way is to consider the relation for the spread-adjusted
return $(p_{\rm bid}(\tau + h)  -   p_{\rm ask}(\tau))/p(\tau + h)$
(written here for a long position). If we now routinely extend the
2-end-point, Eq.(\ref{eq:Pq2p}), we obtain
\begin{equation}
P_q^{\rm ab}(\tau,h)  =
f_q\left(
\left( \frac{  p_{\rm bid}(\tau+h)   -            p_{\rm ask}(\tau)    }{   p(\tau + h)}  \right)
\left( \frac{  p_{\rm bid}(\tau+l_{\rm s})   -    p_{\rm ask}(\tau+h)  }{   p(\tau
+l_{\rm s}) }  \right)
\right)\,.
\label{eq:Pq2pab}
\end{equation}
However this clearly violates, the Dirichlet boundary conditions,
Eq.(\ref{eq:Dir}). The spread itself yields a negligible correction
to the mean values.

\begin{figure}
  \begin{center}
  \includegraphics[width=17cm]{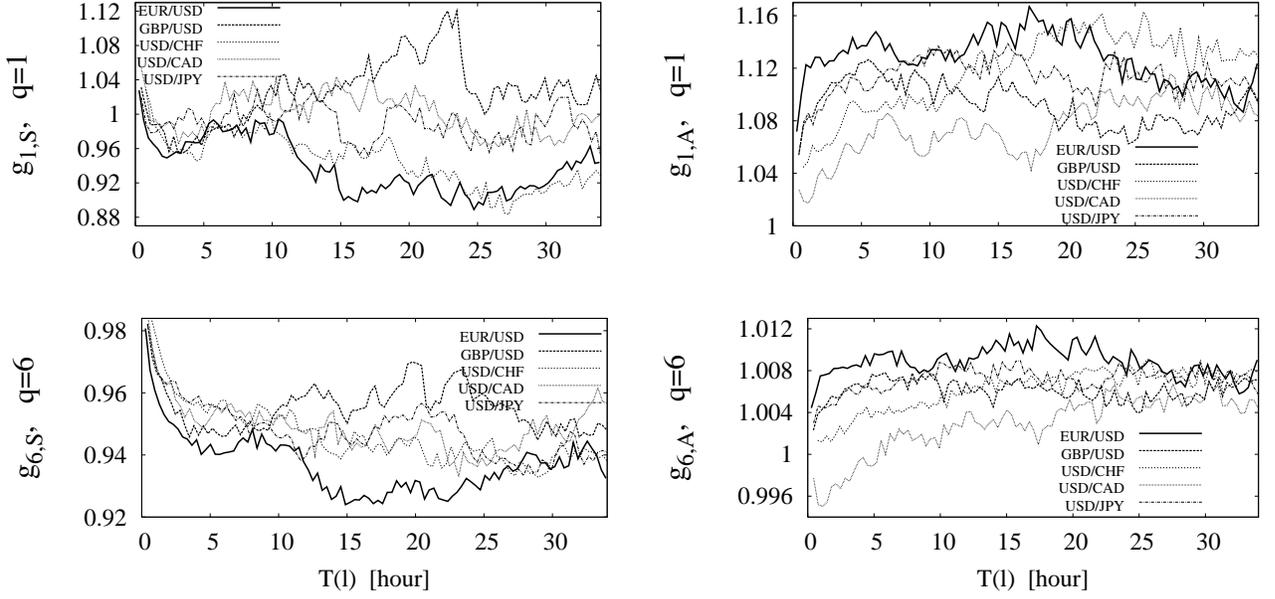}
  \caption{The effect of string polarization on the mean
   characteristics calculated for selected
   currency pairs.
   One might identify a sudden increase of intra-string
   distinguishing features
   for the scales ranging
   from to 3-5 hours.}
  \label{fig:antena}
  \end{center}
  \end{figure}

The boundary conditions can be easily renewed by the subtraction
${\tilde{P}}_q^{\rm ab}(\tau,h)= P_q^{\rm ab}(\tau,h)-P_q^{\rm
ab}(\tau,0)$. However, we show there exists a more fundamental
alternative way which reflects a bid-ask difference and preserves
the Dirichlet boundary conditions. The string states are {\em
polarized} by the instant possibility to place a
successful/unsuccessful buy order. For each $h$ and $Y=A, S$ we
construct the inequality constrained sequence
\begin{eqnarray}
P^{(2),Y}_{(q,+)}(\tau+1,h)=
\left\{\begin{array}{rl}  P^{(2),Y}_q(\tau,h)\,, &
p_{\rm bid}(\tau+l_{\rm s}) \geq  p_{\rm ask}(\tau)\\
P^{(2),Y}_{(q,+)}(\tau,h)\,, &
\mbox{otherwise}
\\
\end{array}
\right.
\end{eqnarray}
and non-buy contributions, respectively,
\begin{eqnarray}
P^{(2),Y}_{(q,-)}(\tau+1,h)
=
\left\{ \begin{array}{ll}
P^{(2),Y}_q(\tau,h)\,,              &         p_{\rm bid}(\tau+l_{\rm s})<p_{\rm ask}(\tau)\\
P^{(2),Y}_{(q,-)}(\tau,h)\,,        &         \mbox{otherwise}\,.
\\
\end{array}
\right.
\end{eqnarray}
In both cases it is supposed that $P^{(2),Y}_q(\tau,h)$ is
calculated according an unconditioned model defined by
Eq.(\ref{eq:Pq2p}). Now, to characterize the arbitrage
opportunities, we introduced the statistical polarization measure in
the form
\begin{equation}
g_{q,Y}=
\Big\langle\frac{
\sum_{h=0}^{l_{\rm s}}
| P^{(2),Y}_{(q,+)}(\tau,h)  -  P^{(2),Y}_{(q,-)}(\tau,h)|}{
\sum_{h=0}^{l_{\rm s}}
| P^{(2),Y}_{(q,+)}(\tau,h)  +  P^{(2),Y}_{(q,-)}(\tau,h)|}
\Big\rangle\,,
\quad  Y= A, S\,\,.
\end{equation}
The results of the extensive study of this measure are
depicted in Fig.\ref{fig:antena}.

\begin{figure}
 \begin{center}
 \includegraphics[width=16cm]{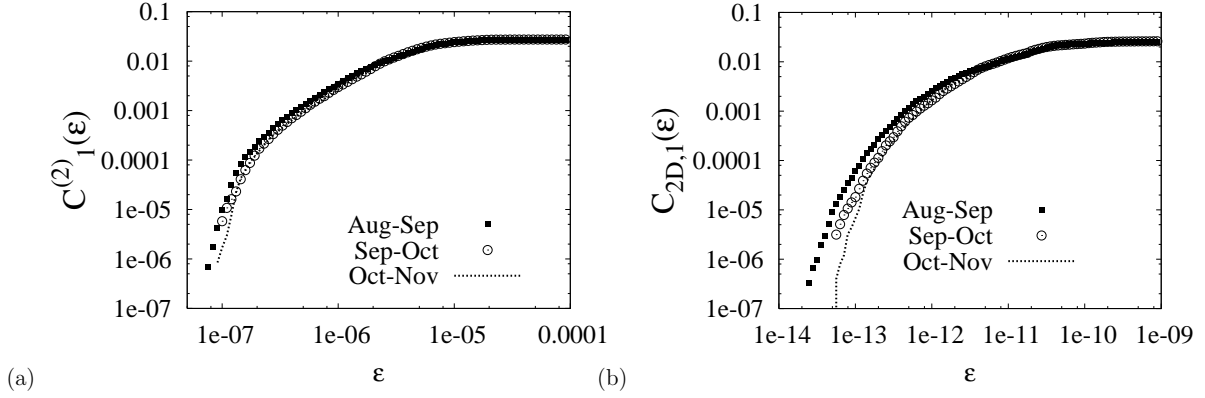}
 \caption{The figure showing the statistics
 of inter-string distances.  In part (a)~the statistics collected and organized
 for the 2-end-point strings by means of the correlation sum (see~\protect{Eq.(\ref{eq:dist1}}))
 for the currency pair GBP/USD and three time periods. As expected, the function $C^{(Y)}_q(\epsilon)$
 monotonically decreases to zero as $\epsilon \rightarrow 0$.
 The index $D_{\rm F}= 1.25 - 1.26$
 varies over the periods.
 The values have been obtained by fitting
 over the estimated {\em universality range}
 $<10^{-7}, 2.10^{-6} >$ of $\epsilon$.  In part (b)~the exponential dependences $C_{{\rm 2D},1}(\epsilon)$
 are obtained after the replacement of
 $d_q^{(Y)}(\tau)$ by $d_{{\rm 2D},q}(\tau)$
 in Eq.(\ref{eq:CYq}). We see that the inter-brane distances
 follow the exponential-type law with the highest variability across
 the short distances.}
 \label{fig:corC}
 \end{center}
 \end{figure}

We continue the characterization of arbitrage opportunities by
defining the(momentum) distance function between the strings as
\begin{equation}
d_q^{(Y)}(\tau) =
\frac{1}{l_{\rm s}+1}
\sum_{h=0}^{l_{\rm s}}
\left| \, P_{(q,+)}^{(Y)}(\tau,h) -  P_{(q,-)}^{(Y)}(\tau,h)
\right|\,.
\label{eq:dist1}
\end{equation}
In this case the statistics of string distances can be characterized
by the customized variant of the well-known model of the {\em
correlation} {\em sum}~\cite{Grass1983,Campbell1997,Bigdeli2009}.
However, the motivation here differs from that given in these
papers, where the intent was to analyze nonlinear relationships. The
correlation sum shows the probability that the states of two strings
or branes are localized within a certain distance. In our view we
adjust the original formula to the string and brane models which, in
addition, reflect the transactions involving profits. We define the
measure
\begin{equation}
C^{(Y)}_q(\epsilon) = \frac{\langle \Theta(\epsilon - d_q^{(Y)}(\tau))\rangle }{
\int {\rm d} \epsilon'\, \langle \Theta(\epsilon' -   d_q^{(Y)}(\tau))\rangle }\,,
\label{eq:CYq}
\end{equation}
where $\epsilon$ is the threshold distance, $\Theta(.)$ is the
Heaviside step function; here $l_{\rm s}$ plays the role of so
called embedding dimension. The key concept surrounding this measure
is the concept of the fractal dimension.  In Fig.(\ref{fig:corC}) we
estimated a specific fractal dimension $D_{\rm F}$  as a slope of
the dependence ${\rm d}\ln C^{(Y)}_q(\epsilon) =  D_{\rm F} {\rm d}
\ln \epsilon$. However, as the figure shows, we have extended our
construction further and extend the concept of distance for 2D
branes as defined in Eq.(\ref{eq:P2Dq}). In a case like this we
suggest generalization
\begin{equation}
d_{{\rm 2D},q}(\tau) =\frac{1}{(l_{\rm s}+1)^2}
\sum_{h_1=0}^{l_{\rm s}} \sum_{h_2=0}^{l_{\rm s}}
\left|
P_{{\rm 2D},(q,+)} (\tau, h_1,h_2) -
P_{{\rm 2D},(q,-)} (\tau, h_1,h_2)
\,
\right|\,.
\label{eq:dist2D}
\end{equation}

\section{Inter-currency study: map onto rotating strings}\label{sec:rotstring}

The incorporating of the mutual relations between the pairs into the
mapping procedure represents a very challenging task. Let us study
trading activity in the $(I,J)$ plane, where $I,J$ stands for
indices of the currency pair described by two 2-end-point strings.
The real time data are used instead of tick by tick (see
sec.\ref{sec:Data}) in order to maintain the consistency of prices
quoted.

\begin{figure}
 \begin{center}
 \includegraphics[width=16cm]{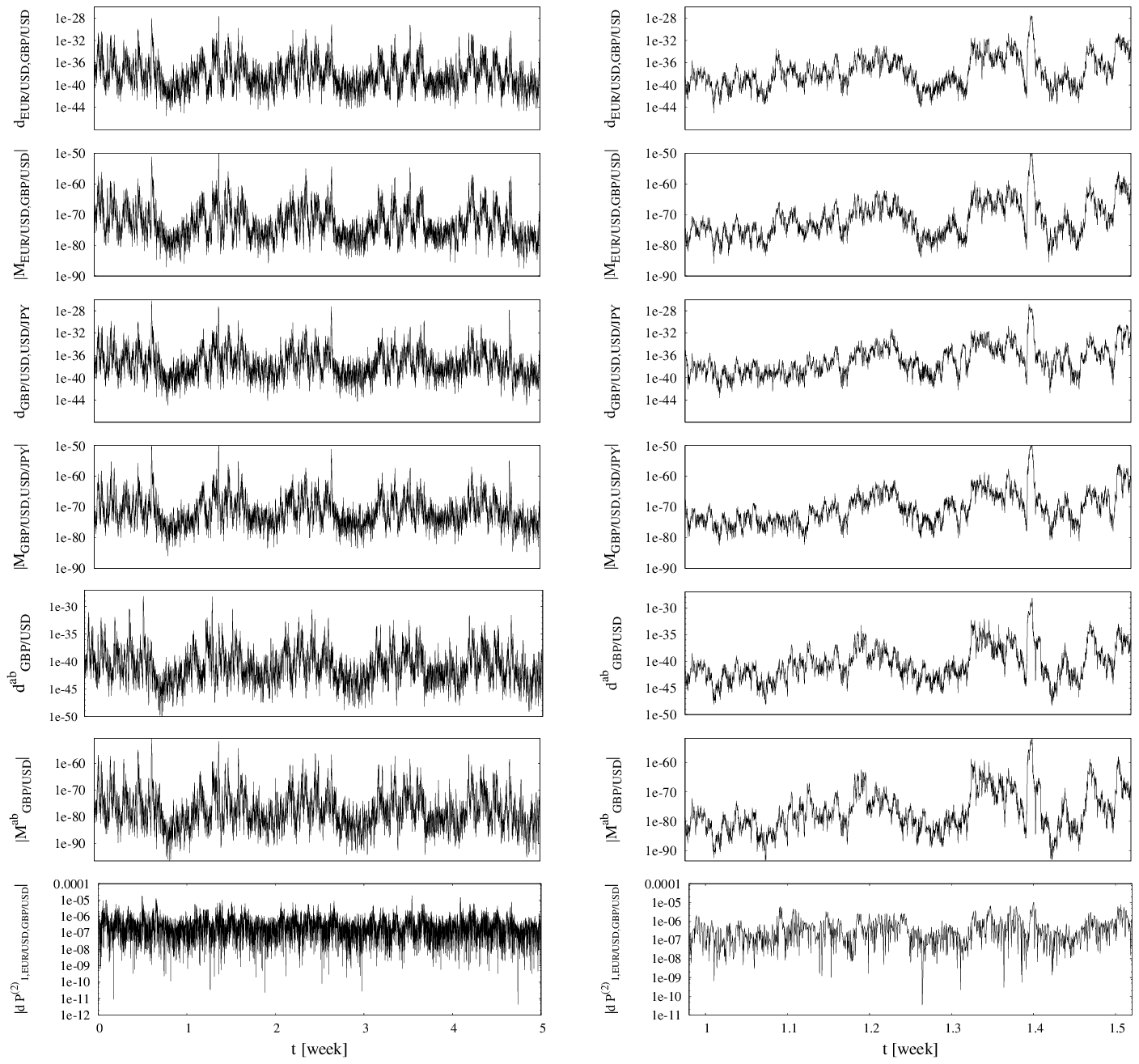}
 \caption{ The inter-string distance and angular momentum
 (see Eqs.(\ref{eq:dis}) and (\ref{eq:mom})) for specified currency pairs.
 The long term outlook compared with a detailed one-week view
 supplemented by the results obtained for
 spread according to Eq.(\ref{eq:disab}) and Eq.(\ref{eq:momab}).
 The results are compared  with the G\mbox{$\hat{a}$}teaux derivative
 (see Eq.(\ref{eq:dP12}), where $p$ corresponds to EUR/USD and $\psi$ to GBP/USD
 pair), and the derivative is evaluated at $h=l_{\rm s}/2$. The calculation is
carried for the string of the 1-hour time length and investigation
of the Aug-Sep period.}
 \label{fig:D12M15}
 \end{center}
 \end{figure}

By continuing examples of the generalized distance concept (see
Eq.(\ref{eq:dist1})) and Eq.(\ref{eq:dist2D}), we can introduce the
inter-currency momentum distance function
\begin{eqnarray}
d_{q,I,J}(t) &=& \frac{1}{l_{\rm s}+1}\,
\sum_{h=0}^{l_{\rm s}}\,
\Big|P^{(2)}_{q,I}(t,h) -  P^{(2)}_{q,J}(t,h) \Big|\,.
\label{eq:dis}
\end{eqnarray}
At higher dimension, it is tempting to deal with angular momentum
\begin{eqnarray}
M_{q,I,J}(t)  =
\sum_{h=0}^{l_{\rm s}} \, \Big[\, P^{(2)}_{q,I}(t,h) X^{(2)}_{q,J}(t,h)  -
P^{(2)}_{q,J}(t,h) X^{(2)}_{q,I}(t,h)\,\Big]\,.
\label{eq:mom}
\end{eqnarray}
The momentum calculation can be interpreted as a measure of the
rotational information flows between the currency pairs. In
Fig.(\ref{fig:D12M15}) we present the results achieved for two
alternatives  $(I, J) = (\mbox{EUR/USD}, \mbox{GBP/USD})$, and
$(I,J)=(\mbox{GBP/USD},\mbox{USD/JPY})$. We see that calculation of
$D_{q,I,J}$ and $M_{q,I,J}$ yields oscillatory (i.e. seasonal)
behavior. Simultaneously, the concept of distance and moment has
been extended to analyze the impact of spread. Analogously, as in
the previous cases, the distance between the ask and bid strings may
be defined
\begin{eqnarray}
d_{q,I}^{\rm ab}(t) &=& \frac{1}{l_{\rm s}+1}\,
\sum_{h=0}^l\,
\Big|
P^{(2)}_{q,{\rm ask}}(t,h) -P^{(2)}_{q,{\rm bid}}(t,h) \Big|\,,
\label{eq:disab}
\\
M_{q,I}^{\rm ab}(t)  &=&  \sum_{h=0}^{l_{\rm s}} \,
\Big[\,
P^{(2)}_{q,{\rm ask}}(t,h)
X^{(2)}_{q,{\rm bid}}(t,h)  -
P^{(2)}_{q,{\rm bid}}(t,h)
X^{(2)}_{q,{\rm ask}}(t,h)\Big]\,.
\label{eq:momab}
\end{eqnarray}
Here
\begin{eqnarray}
P^{(2)}_{q,{\rm ask}} \equiv  P^{(2)}_q |_{p \rightarrow p_{\rm ask}}
\,,
\quad
P^{(2)}_{q,{\rm bid}} \equiv  P^{(2)}_q |_{p \rightarrow p_{\rm bid}}
\end{eqnarray}
are obtained by substituting expressions above in
Eq.(\ref{eq:Pq2p}). With the help of Eq.(\ref{eq:Xq2}) and
$P^{(2)}_{q,{\rm ask}}$, $P^{(2)}_{q,{\rm bid}}$ we construct
iteratively $X^{(2)}_{q,{\rm ask}}$ and $X^{(2)}_{q,{\rm bid}}$. As
one can see from Fig.(\ref{fig:D12M15}), the differences measured in
terms of $M_{q,I,J}(t)$ are very subtle. There is evidence of
intercoupling of the spread and currency dynamics. The fundamental
role in the string theory is played by the {\em Regge} {\em slope}
{\em parameter} $\alpha'$ (or inverse string tension). This has a
proper analogy with our approach where we introduced a slope in
terms of the angular momentum
\begin{equation}
\alpha_{q,I,J}'  =  \frac{ <| M_{q,I,J} | >}{ 2\pi\, l_{\rm s}^2}\,.
\label{eq:MqIJ}
\end{equation}
For the $l_{\rm s}=1 \mbox{hour}$ string pair we obtained
$\alpha_{6, \mbox{\tiny EUR/USD, GBP/USD}}' $ $ = 5.07 . 10^{-54}$
$(2\pi)^{-1} $ $ \mbox{hour}^{-2}$, $\alpha_{6, \mbox{\tiny GBP/USD,
USD/JPY}}' $ $ = 1.55 . 10^{-53}$ $(2\pi)^{-1} $ $
\mbox{hour}^{-2}$, $\alpha_{6, \mbox{\tiny USD/JPY, EUR/USD}}' $ $ =
1.34 . 10^{-53}$ $(2\pi)^{-1} $ $ \mbox{hour}^{-2}$ much larger than
the spread ${\alpha_{6, \mbox{\tiny GBP/USD}}^{,\rm ab}} = 1.16
\cdot 10^{-55} \,(2\pi)^{-1} \mbox{hour}^{-2}$. However, it is worth
noting that relation, Eq.(\ref{eq:MqIJ}), should be understood as an
estimate since there is no statistical mean of the type $<|
M_{\ldots} | >$ in the original specification.  The problem of
estimation of the {\em slope parameter} arises from the fact that in
the original model nonaveraged angular momentum is divided by the
square of the mass instead of $l_{\rm s}^2$.  Herein, we have no
idea how to measure the mass of the string, or how to verify the
fact that the string mass is proportional to $l_{\rm s}$.

\section{Differentials of string map}\label{sec:Gateaux}

G\^{a}teaux derivative is a generalization of the concept of a
directional derivative in the differential calculus. In our study
the concept can be viewed as a systematic way in the generation of
more structured maps expressing more information about the structure
of data we deal with. Given the string map $P(.)$, the m-th {\em
G\^{a}teaux derivative} of $P(.)$ in the "direction" of $\psi(.)$
(unspecified yet series) is defined as follows:
\begin{eqnarray}
&& {\rm d}^m P(\{p\}; \{\psi\} )(\tau,h) =
\frac{{\rm d}^m}{{\rm d} \epsilon^m} P(\{p(\tau,h) +
\epsilon \psi(\tau,h) \}){\Big|}_{\epsilon\rightarrow 0} \,\,.
\end{eqnarray}
For $q=1$ the calculation gives
\begin{equation}
{\rm d} P_1^{(1)}(\{p\};\{\psi\} )(\tau,h) =
\frac{1}{p(\tau+h)} \left[\frac{ p(\tau)  \psi(\tau +h)}{p(\tau+h)}  - \psi(t) \right]
\end{equation}
and
\begin{eqnarray}
&&
{\rm d} P_1^{(2)}(\{p\}; \{\psi\} )(\tau,h)
=\psi(\tau)\left(\frac{1}{p(\tau + l_{\rm s})} - \frac{1}{p(\tau+h)}\right)
\label{eq:dP12}
\\
&+&
\psi(\tau+h)\left( \frac{p(\tau)}{p^2(\tau+h)} - \frac{1}{p(\tau + l_{\rm s})}   \right)
+
\frac{\psi(\tau  +  l_{\rm s})}{p^2(\tau+l_{\rm s})}   \left( p(\tau+h) -  p(\tau) \right)
\nonumber\,.
\end{eqnarray}
By going to the
second order we obtained
\begin{equation}
{\rm d}^2 P_1^{(2)}(\{p\}; \{\psi\} )(\tau,h) = \frac{2\psi(\tau+h)}{p^2(\tau+h)}\,
\left[\,
\psi(\tau)- \frac{p(\tau) \psi(\tau+h)}{ p(\tau+h) }\, \right]
\end{equation}
and
\begin{eqnarray}
&&
{\rm d}^2 P_1^{(2)}(\{p\}; \{\psi\})(\tau,h) = \frac{2\psi(\tau+l_{\rm s})}{p^2(\tau+l_{\rm s})}
\left[\psi(\tau+h)-\psi(\tau)\right]
\\
&+&
\frac{2 \psi(\tau+h)}{p^2(\tau+h)} \left[\psi(\tau) - \frac{p(\tau) \psi(\tau+h)}{p(\tau+h)}\right]+
\frac{2 \psi^2(\tau+l_{\rm s})}{p^3(\tau+l_{\rm s})} \left[p(\tau)-p(\tau+h) \right]\,.
\nonumber
\end{eqnarray}
Surprisingly, the generalized differentiation generates maps which
satisfy the Dirichlet boundary conditions
\begin{eqnarray}
&&  {\rm d}^m P_1^{(1)}(\{p\}; \{\psi\} )(\tau,0) =  0\,, \qquad m=1,2;
\\
&&  {\rm d}^m P_1^{(2)}(\{p\}; \{\psi\} )(\tau, 0)
=   {\rm d}^m P_1^{(2)}(\{p\}; \{\psi\})(\tau, l_{\rm s}) = 0\,.
\end{eqnarray}
Many alternative ways exist to exploit the models with the auxiliary
field $\psi(.)$. The field can be related to, e.g., (i)~models which
place emphasis on the currency margins determined by some adaption
process; (ii)~on the spread in a style of sec.\ref{sec:polar} with
$\psi(\tau)  =  p_{\rm bid}(\tau) - p_{\rm ask}(\tau-l_{\rm s})$.
(iii)~The benchmark setting represents $\psi=1$, (iv)~the periodic
function $\psi(\tau)$ can model the action of the compact. In this
intuition supporting case, one can see that the generalized
derivative modifies the original map as follows:
\begin{eqnarray}
{\rm d} P_1^{(1)}(\{p\};\{\psi\})(\tau,h)|_{\psi=1}
&=&  - \frac{P_1^{(1)}(\tau,h)}{p(\tau+h)}\,,
\\
{\rm d} P_1^{(2)}(\{p\};\{\psi\})(\tau,h)|_{\psi=1}
&=&
\left( \frac{1}{p(\tau+h)} + \frac{1}{p(\tau+l_{\rm s})}
\right)\, P_1^{(2)}(\tau,h)\,.
\end{eqnarray}
In Fig.\ref{fig:D12M15}, we present the idea, where $p$, $\psi$ are
represented by two currencies, their mutual influence is studied
within the first-order differential model described by
Eq.(\ref{eq:dP12}).

\section{Conclusions}

We shown that the string theory may motivate the adoption of the
nonlinear techniques of the data analysis with a minimum impact of
justification parameters. The numerical study recovered interesting
fundamental statistical properties of the maps from the data onto
string-like objects. The remarkable deviations from the features
known under the notion of the efficiently organized market have been
observed, namely, for high values of the deformation parameter $q$.

The main point here is that the string map gives a geometric
interpretation of the information value of the data. The model of
the string allows one to manipulate with the information stored
along several extra dimensions. We started from the theory of the
1-end-point and 2-end-point string, where we distinguished between
the symmetric and antisymmetric variants of the maps.   In this
context, it should be emphasized that duality is a peculiar property
of the suggested maps, not data alone.  The numerical analysis of
the intra-string statistics was supplied qualitatively by the toy
models of the maps of the exponential and periodic data inputs. Most
of the numerical investigations have been obtained for the open
topology; however, we described briefly the ways to partial
compactification. The data structures can also be mapped by means of
the curled dimension which arises as a sum of periodic data
contributions. The idea of the compactified strings can be realized
as well by the application of the inverse Fourier transform of the
original signal. The interesting and also challenging 
task represents finding of link between string map and log-periodic 
behaviour of speculative bubbles of the stock market 
indices~\cite{Zhou2003,Zhou2004}. It would be also interesting 
to examine R/S analysis of the Hurst exponents~\cite{Mandelbrot1971,Granero2008} 
for the case of finite strings instead of the usual point prices. 

The study of string averages exhibited occurrences of the anomalies
at the time scales proportional to the string length. We showed that
global and common market timescales can be extracted by looking at
the changes in the currencies. The extensions of the string models
of branes including ask/bid spread were discussed. We studied the
relationship between the arbitrage opportunities and string
statistics. We showed that extraction of the valuable information
about the arbitrage opportunities on given currency could be studied
by means of the correlation sum which reflected the details of the
occupancy of phase-space by differently polarized strings and
branes. In addition, we presented several physics and geometry
motivated methods of analysis of the coupling between the currency
pairs. The results led us to believe that our ideas and methodology
can contribute to the solution of the problem of the robust
portfolio selection. As we have seen, the complex multi-string
structures produced by the generalized derivatives of strings cannot
be easily grasped by the intuitive principles.  We believe, the
method affords potential to be used in the practical applications,
where arbitrage selection bias should be taken into account.

\end{document}